\documentclass[12pt,thmsa,a4paper,notitlepage]{article}
\usepackage{amssymb}

\usepackage{sw20lart}

\input{tcilatex}
\begin{document}

\author{S.Balaska\thanks{%
e-mail : balaska@univ-oran.dz} and K. Demmouche\thanks{%
e-mail : k\_demmouche@yahoo.fr } \and \textit{Laboratoire de physique
th\'{e}orique d'Oran. } \and \textit{D\'{e}partement de physique.
Universit\'{e} d'Oran Es-S\'{e}nia .} \and \textit{31100 Es-S\'{e}nia.
ALGERIA}}
\title{The correlation functions of the $(D_{4},A_{6})$\ conformal model}
\maketitle

\begin{abstract}
In this work, we exploit the operator content of the $(D_{4},A_{6})$\
conformal algebra. By constructing a $Z_{2}$-invariants fusion rules of a
chosen subalgebra and by resolving the bootstrap equations consistent with
these rules, we determine the structure constants of the subalgebra.

\medskip \medskip

\end{abstract}

\section{Introduction}

The biggest advantage in studying two dimensional conformal field theories,
is the presence of an infinite amount of symmetries. This fact is very
helpful to compute the correlators of the fields present in the theory.

The first approach proposed to reach this goal is the bootstrap approach 
\cite{PBZ}. Using the coulomb gas formulation, Dotsenko and Fateev \medskip
have solved, in \cite{Dotsenko}, the bootstrap equations and determined the
conformal blocks as well as the structure constants of the operator algebras
associated with the minimal models. These models are shown later to form the
A-series of the A-D-E classification obtained in \cite{capelli}. They
involve only spinless fields.

In the case of the models of the D-series, involving also spin fields, many
approaches have been proposed for the determination of the structure
constants and the correlators (see for example \cite{petkova}, \cite{Mccabe1}%
, \cite{Mccabe2}, \cite{runkel1} and \cite{rida}).

In \cite{Mccabe1} the authors have determined the structure constants
appearing in the three point correlation functions for the particular case
of the $(A_{4},D_{4})$-model or what they call the D$_{5}$-model. They have
used the well known coulomb gas technique adapted to the cases where the
correlators involve spin fields. Their results depend on the choice of the $%
Z_{2}$-symmetry used to separate the two copies of certain fields present in
the theory. Among the four different possibilities, they took only one by
imposing the consistency of the results with the bootstrap equations. Their
choice has led to the vanishing of one of the structure constants. This fact
was not predicted by the non-chiral $Z_{2}$\ invariant fusion rules, but at
the same time it does not disagree with a non-zero $Z_{3}$-invariant 3-pt
function of extended fields \cite{Di fran}. The implementation of such
discrete symmetries in non diagonal models is discussed in \cite{ruelle}%
\textbf{. }

In our present work, we apply exactly the same method, as in \cite{Mccabe1},
to determine the structure constants of the $(D_{4},A_{6})$-model. The only
difference is that we consider a different $Z_{2}$-transformation. Our
results are consistent with the bootstrap equations and also with the
non-chiral $Z_{2}$\ invariant fusion rules.

The paper is organised as follows. In section 2, we start with the operator
content of our model. After that, we write the fusion rules and exhibit the
different choices for the \medskip $Z_{2}$-symmetries. In section 3 , we
derive the desired $Z_{2}$-invariant fusion rules of our operator algebra.
In section 4 and 5, we give the essential of the method used to determine
the structure constants. In section 6, we apply it to the $(D_{4},A_{6})$%
-model and give the conclusion.

\section{The conformal $(D_{4},A_{6})\ $Model}

The partition functions of unitary conformal field theories defined on the
torus were classified in three infinite series : the A-D-E-series \cite
{capelli}. In a generic case, for a modular invariant model corresponding to
the minimal one $\mathcal{M}(p=m+1,p^{\prime }=m)$ with 
\begin{equation}
c=1-\frac{6}{m(m+1)}<1\ \text{ \qquad ,}\qquad m=2,3,4...,  \label{eq2-1}
\end{equation}
the partition function is written as a sesquilinear form in the characters $%
\chi _{h\ \ }(\overline{\chi }_{\overline{h}})$ of the representations of
the left (right) Virasoro algebra generated by the primary fields $\phi
_{h}(z)$ ($\phi _{\overline{h}}(\overline{z}))$

\begin{equation}
Z=\sum_{h,\overline{h}\in E(p,p^{\prime })}N_{h\overline{,h}\ \ }\chi _{h\ \
}\overline{\chi }_{\overline{h}}\ \ ,  \label{eq2-2}
\end{equation}
where $E(p,p^{\prime })$ is the set of the conformal weights given by the
Kac formula 
\begin{eqnarray}
h &=&h_{r,s}=\frac{[r(m+1)-sm]^{2}-1}{4m(m+1)}  \nonumber \\
1 &\leqslant &r\leqslant m-1\ \ \qquad ;\qquad 1\leqslant s\leqslant m\qquad
\label{eq2-3}
\end{eqnarray}
and having the symmetry property 
\[
h_{r,s}=h_{p^{\prime }-r,p-s}. 
\]
The nonnegative integers $N_{h\overline{,h}\ \ }$denote the multiplicities
of occurrence of the corresponding left-right representation modules (the
identity operator $\Phi _{(1,1),(1,1)}$ being nondegenerate, makes $%
N_{h_{1,1},\overline{h}_{1,1}}=N_{0,0}=1$ ).

The characters $\chi _{h\ \ }$and $\overline{\chi }_{\overline{h}}$ are
shown to form a unitary representation of the modular group $PSL(2,Z).$ So
if we note $N$ the matrix whose elements are the integers $N_{h\overline{,h}%
\ \ }$, the problem of finding all possible modular invariant forms of
equation (\ref{eq2-2}) is reduced to the problem of finding all the matrices 
$N$ commuting with those representing the generators $S$ and $T$ of the
modular group (in \cite{Di fran} one can find the explicit elements of the
matrices $S$ and $T$).

For a given minimal model with $(m\geq 5)\ $there is always at least two
solutions to this problem. A trivial one with 
\[
N_{h\overline{,h}\ \ }=\delta _{h\overline{,h}\ \ }, 
\]
corresponds to the diagonal modular invariants models of the A-series 
\begin{equation}
Z_{(A_{p^{\prime }-1},A_{p-1})}=\sum_{(r,s)\in E(p,p^{\prime })}\left| \chi
_{r,s}\right| ^{2},  \label{eq 2-4}
\end{equation}
here $\chi _{r,s}=\chi _{h_{r,s}\text{ }}$. The second solution (less
trivial) gives the modular invariants models of the D-series. Those
associated with the minimal models with $p^{\prime }=m=2(2n+1)$ (to which
belongs our model) are 
\begin{equation}
Z_{(D_{p^{\prime }/2+1},A_{p-1})}=\frac{1}{4}\sum\Sb (r,s)\in E(p,p^{\prime
})  \\ r\ \ odd  \endSb \left| \chi _{r,s}+\chi _{p^{\prime }-r,s}\right|
^{2}.  \label{eq 2-5}
\end{equation}
Then for the model with $m=6$, associated to the minimal one $\mathcal{M}%
(7,6)$ we have 
\begin{equation}
Z_{(D_{4},A_{6})}=\sum_{s=1,2,3}\left| \chi _{1,s}+\chi _{5,s}\right|
^{2}+2\left| \chi _{3,s}\right| ^{2}.  \label{eq 2-6}
\end{equation}
The operator content of this model, resumed in table 1, can be read off
directly from the last equation. We note that not all the fields of the $%
\mathcal{M}(7,6)\ $minimal model are present , the existence of spinless as
well as spin left-right combinations and finally the presence of two copies
for each of the spinless combinations $\Phi _{(3,s|3,s)}$\medskip $=\phi
_{(3,s)}\otimes \overline{\phi }_{(3,s)}$ \ for $s=1,2,3$. \medskip A third
solution exist for some exceptional minimal models , it gives the modular
invariants of the E-series (see \cite{Di fran} for a complete review on the
conformal field theory).

\ \ \ \ \ \ \ \ \ \ \ \ \ \ \ \ \ \ \ \ \ \ \ \ \ \ \ \ \ 
\begin{table}[tbp] \centering%
\begin{tabular}{|c|c|c|c|}
\hline\hline
\multicolumn{1}{||c|}{$\Phi _{(r,s),(r^{\prime },s^{\prime })}$} & 
\multicolumn{1}{||c|}{$(h,\overline{\mathit{h}})$} & \multicolumn{1}{||c|}{$%
s=h-\overline{h}$} & \multicolumn{1}{||c||}{{\small N}$_{h,\overline{h}}$}
\\ \hline\hline
$\Phi _{(1,1),(1,1)}$ & $(0,0)$ & 0 & 1 \\ \hline
$\Phi _{(3,1),(3,1)}$ & $(\frac{4}{3},\frac{4}{3})$ & 0 & 2 \\ \hline
$\Phi _{(3,2),(3,2)}$ & $(\frac{10}{21},\frac{10}{21})$ & 0 & 2 \\ \hline
$\Phi _{(3,3),(3,3)}$ & $(\frac{1}{21},\frac{1}{21})$ & 0 & 2 \\ \hline
$\Phi _{(5,1),(5,1)}$ & $(5,5)$ & 0 & 1 \\ \hline
$\Phi _{(1,5),(1,5)}$ & $(\frac{22}{7},\frac{22}{7})$ & 0 & 1 \\ \hline
$\Phi _{(1,4),(1,4)}$ & $(\frac{12}{7},\frac{12}{7})$ & 0 & 1 \\ \hline
$\Phi _{(1,3),(1,3)}$ & $(\frac{5}{7},\frac{5}{7})$ & 0 & 1 \\ \hline
$\Phi _{(1,2),(1,2)}$ & $(\frac{1}{7},\frac{1}{7})$ & 0 & 1 \\ \hline
$\Phi _{(1,1),(5,1)}$ & $(0,5)$ & -5 & 1 \\ \hline
$\Phi _{(5,1),(1,1)}$ & $(5,0)$ & 5 & 1 \\ \hline
$\Phi _{(1,4),(1,3)}$ & $(\frac{12}{7},\frac{5}{7})$ & 1 & 1 \\ \hline
$\Phi _{(1,3),(1,4)}$ & $(\frac{5}{7},\frac{12}{7})$ & -1 & 1 \\ \hline
$\Phi _{(1,5),(1,2)}$ & $(\frac{22}{7},\frac{1}{7})$ & 3 & 1 \\ \hline
$\Phi _{(1,2),(1,5)}$ & $(\frac{1}{7},\frac{22}{7})$ & -3 & 1 \\ \hline
\end{tabular}
\caption{The operator content of the D$_{6}$ model\label{key}}%
\end{table}%

\section{The fusion rules}

The set of the primaries and their descendants, of a given minimal model,
form a closed algebra with respect to the fusion operation. This is in
general an operation among the operators forming representations of chiral
algebras inherited from the operator product algebra. In our case and when
considering only the ''chiral part'' of the conformal theory, i.e only the
holomorphic part (or in the same manner only the anti-holomorphic part) the
fusion between two conformal families is written

\begin{equation}
\left[ \phi _{r,s}\right] \times \left[ \phi _{r^{\prime },s^{\prime
}}\right] =\mathcal{N}_{(r,s),(r^{\prime },s^{\prime })}^{(r^{\prime \prime
},s^{\prime \prime })}\left[ \phi _{r^{\prime \prime },s^{\prime \prime
}}\right]  \label{eq 2-7}
\end{equation}
\medskip where the fusion coefficients $\ \{\mathcal{N}_{i,j}^{k}\}$ take
the values $1$ or $0$ to indicate if at least one member of the conformal
family $\left[ \phi _{r^{\prime \prime },s^{\prime \prime }}\right] $ is
present in the fusion of the two families $\left[ \phi _{r,s}\right] $ and $%
\left[ \phi _{r^{\prime },s^{\prime }}\right] .$ This looks similar to the
decomposition of the usual tensor product of representations in terms of the
irrep. There is a formula, due to Verlinde \cite{verlinde}, expressing these
coefficients for the minimal models in terms of the elements of the unitary
matrix $S$ of the modular group 
\begin{equation}
\mathcal{N}_{(r,s),(r^{\prime },s^{\prime })}^{(r^{\prime \prime },s^{\prime
\prime })}\medskip =\sum_{(k,l)\in E(p,p^{\prime })}\frac{S_{(r,s),(k,l)}\
S_{(r^{\prime },s^{\prime }),(k,l)}\ S_{(k,l),(r^{\prime \prime },s^{\prime
\prime })}}{S_{(1,1),(k,l)}}  \label{eq 2-8}
\end{equation}
with 
\begin{equation}
S_{(r,s),(r^{\prime },s^{\prime })}\medskip =2\sqrt{\frac{2}{pp^{\prime }}}%
(-)^{1+sr^{\prime }+rs^{\prime }}\sin (\pi \frac{p}{p^{\prime }}rr^{\prime
})\sin (\pi \frac{p^{\prime }}{p}ss^{\prime })  \label{eq 2-8'}
\end{equation}
On the operator algebra, the fusion operation is realized by the short
distance expansion of the operators product 
\begin{equation}
\Phi _{h_{1},\overline{h}_{1}}(z,\overline{z})\Phi _{h_{2},\overline{h}%
_{2}}(0,0)=\sum_{h,\overline{h}}\frac{\widetilde{C}_{(h_{1},\overline{h}%
_{1}),(h_{2},\overline{h}_{2}),(h,\overline{h})}}{z^{h_{1}+h_{2}-h}\ \ 
\overline{z}^{\overline{h}_{1}+\overline{h}_{2}-\overline{h}}}\Phi _{h,%
\overline{h}}(0,0)\left[ 1+\mathcal{O}(z,\overline{z})\right]  \label{eq 2-9}
\end{equation}
where the contribution of the descendants is contained in $\mathcal{O}(z,%
\overline{z}).$ The coefficients appearing in the expansion are the
structure constants of the operator product algebra. They are of very
interest because as we will see later they appear also in the 3-point
correlation functions.

\medskip For our model , and if we restrict ourselves to the subalgebra
containing the operators : $\Phi _{(1,1),(1,1)}$ , $\Phi _{(1,1),(5,1)}$ , $%
\Phi _{(5,1),(1,1)}$ , $\Phi _{(3,1),(3,1)}$ and $\Phi _{(5,1),(5,1)}$ , we
have for the chiral fusion rules , i.e the fusion rules between the
holomorphic parts only or the antiholomorphic ones 
\begin{eqnarray}
\left[ \phi _{3,1}\right] \times \left[ \phi _{3,1}\right] &=&\left[ \phi
_{1,1}\right] +\left[ \phi _{3,1}\right] +\left[ \phi _{5,1}\right] 
\nonumber \\
\left[ \phi _{5,1}\right] \times \left[ \phi _{5,1}\right] &=&\left[ \phi
_{1,1}\right]  \label{eq2-10} \\
\left[ \phi _{3,1}\right] \times \left[ \phi _{5,1}\right] &=&\left[ \phi
_{3,1}\right]  \nonumber
\end{eqnarray}

\medskip Now combining both the left and right part we obtain the non-chiral
fusion rules 
\begin{eqnarray}
\lbrack \Phi _{(3,1),(3,1)}]\times [\Phi _{(3,1),(3,1)}] &=&[\Phi
_{(1,1),(1,1)}]+[\Phi _{(1,1),(5,1)}]+[\Phi _{(5,1),(1,1)}]  \nonumber \\
&&+[\Phi _{(5,1),(5,1)}]+[\Phi _{(3,1),(3,1)}]  \nonumber \\
\lbrack \Phi _{(5,1),(5,1)}]\times [\Phi _{(5,1),(5,1)}] &=&[\Phi
_{(1,1),(1,1)}]  \nonumber \\
\lbrack \Phi _{(3,1),(3,1)}]\times [\Phi _{(5,1),(5,1)}] &=&[\Phi
_{(3,1),(3,1)}]  \nonumber \\
\lbrack \Phi _{(5,1),(1,1)}]\times [\Phi _{(5,1),(1,1)}] &=&[\Phi
_{(1,1),(1,1)}]  \nonumber \\
\lbrack \Phi _{(5,1),(1,1)}]\times [\Phi _{(1,1),(5,1)}] &=&[\Phi
_{(5,1),(5,1)}]  \label{eq2-11} \\
\lbrack \Phi _{(3,1),(3,1)}]\times [\Phi _{(5,1),(1,1)}] &=&[\Phi
_{(3,1),(3,1)}]  \nonumber \\
\lbrack \Phi _{(3,1),(3,1)}]\times [\Phi _{(1,1),(5,1)}] &=&[\Phi
_{(3,1),(3,1)}]  \nonumber \\
\lbrack \Phi _{(1,1),(5,1)}]\times [\Phi _{(1,1),(5,1)}] &=&[\Phi
_{(1,1),(1,1)}]  \nonumber \\
\lbrack \Phi _{(5,1),(5,1)}]\times [\Phi _{(1,1),(5,1)}] &=&[\Phi
_{(5,1),(1,1)}]  \nonumber \\
\lbrack \Phi _{(5,1),(5,1)}]\times [\Phi _{(5,1),(1,1)}] &=&[\Phi
_{(1,1),(5,1)}]  \nonumber
\end{eqnarray}
To be able to distinguish between the two copies of the field $\Phi
_{(3,1),(3,1)}$ , both present in our subalgebra, one introduces a $Z_{2}$%
-symmetry which gives opposite parities to the two copies. The action of
this transformation on the rest of the fields will be fixed by imposing a
consistency condition on the fusion rules. Let's note $\Phi ^{+}$ and $\Phi
^{-}$ the two copies and take the transformation

\begin{equation}
\Phi ^{\pm }\longrightarrow \pm \Phi ^{\pm }  \label{Z2-symm}
\end{equation}
It's now easy to verify that the fusion 
\[
\lbrack \Phi _{(5,1),(1,1)}]\times [\Phi _{(1,1),(5,1)}]=[\Phi
_{(5,1),(5,1)}] 
\]
is preserved under the four following distinct transformations :

\begin{enumerate}
\item[\textbf{T}$_{1}$]  \medskip $\qquad \Phi _{(5,1),(5,1)}\longrightarrow
+\Phi _{(5,1),(5,1)}\ \ \ \ \ ;\ \ \ \ \ \Phi _{(1,1),(5,1)}\longrightarrow
+\Phi _{(1,1),(5,1)}\ \ \ \ \ \ \ ;\ \ \ \ \ \ \ $

$\ \ \ \ \ \ \ \ \ \ \ \ \ \ \ \ \ \ \ \ \ \ \ \ \ \ \Phi
_{(5,1),(1,1)}\longrightarrow +\Phi _{(5,1),(1,1)}$

\item[\textbf{T}$_{2}$]  $\qquad \Phi _{(5,1),(5,1)}\longrightarrow +\Phi
_{(5,1),(5,1)}\ \ \ \ \ ;\ \ \ \ \ \Phi _{(1,1),(5,1)}\longrightarrow -\Phi
_{(1,1),(5,1)}\ \ \ \ \ \ \ ;\ \ $

$\ \ \ \ \ \ \ \ \ \ \ \ \ \ \ \ \ \ \ \ \ \ \ \ \ \Phi
_{(5,1),(1,1)}\longrightarrow -\Phi _{(5,1),(1,1)}$

\item[\textbf{T}$_{3}$]  $\qquad \Phi _{(5,1),(5,1)}\longrightarrow -\Phi
_{(5,1),(5,1)}\ \ \ \ \ ;\ \ \ \ \ \Phi _{(1,1),(5,1)}\longrightarrow -\Phi
_{(1,1),(5,1)}\ \ \ \ \ \ \ ;\ \ $

$\ \ \ \ \ \ \ \ \ \ \ \ \ \ \ \ \ \ \ \ \ \ \ \ \ \Phi
_{(5,1),(1,1)}\longrightarrow +\Phi _{(5,1),(1,1)}$

\item[\textbf{T}$_{4}$]  $\qquad \Phi _{(5,1),(5,1)}\longrightarrow -\Phi
_{(5,1),(5,1)}\ \ \ \ \ ;\ \ \ \ \ \Phi _{(1,1),(5,1)}\longrightarrow +\Phi
_{(1,1),(5,1)}\ \ \ \ \ \ \ ;\ \ \ \ $

$\ \ \ \ \ \ \ \ \ \ \ \ \ \ \ \ \ \ \ \ \ \ \ \ \ \Phi
_{(5,1),(1,1)}\longrightarrow -\Phi _{(5,1),(1,1)}$
\end{enumerate}

One can then choose four different $Z_{2}$-symmetries. They lead to four
different operator algebras, i.e the fusion rules are contrained differently
in each case. But as we will see later, only the results for the structure
constants obtained from the possibility \textbf{T}$_{2}$ are in total
concordance with the fusion rules and the bootstrap equations. This
transformation was already used in \cite{rida} for the $D_{odd}$ models.
This choice is different from that taken in \cite{Mccabe1}, where the
possibility \textbf{T}$_{1}$ was preferred.

With the possibility \textbf{T}$_{2}$ the fusion rules (\ref{eq2-11}) become
:

\begin{eqnarray}
\lbrack \Phi ^{\pm }]\times [\Phi ^{\pm }] &=&[\Phi _{(1,1),(1,1)}]+[\Phi
_{(5,1),(5,1)}]+[\Phi ^{+}]  \nonumber \\
\lbrack \Phi ^{+}]\times [\Phi ^{-}] &=&[\Phi ^{-}]+[\Phi
_{(1,1),(5,1)}]+[\Phi _{(5,1),(1,1)}]  \nonumber \\
\lbrack \Phi _{(5,1),(5,1)}]\times [\Phi _{(5,1),(5,1)}] &=&[\Phi
_{(1,1),(1,1)}]  \nonumber \\
\lbrack \Phi ^{\pm }]\times [\Phi _{(5,1),(5,1)}] &=&[\Phi ^{\pm }] 
\nonumber \\
\lbrack \Phi _{(5,1),(1,1)}]\times [\Phi _{(5,1),(1,1)}] &=&[\Phi
_{(1,1),(1,1)}]  \nonumber \\
\lbrack \Phi _{(5,1),(1,1)}]\times [\Phi _{(1,1),(5,1)}] &=&[\Phi
_{(5,1),(5,1)}]  \label{eq2-13} \\
\lbrack \Phi ^{\pm }]\times [\Phi _{(5,1),(1,1)}] &=&[\Phi ^{\mp }] 
\nonumber \\
\lbrack \Phi ^{\pm }]\times [\Phi _{(1,1),(5,1)}] &=&[\Phi ^{\mp }] 
\nonumber \\
\lbrack \Phi _{(1,1),(5,1)}]\times [\Phi _{(1,1),(5,1)}] &=&[\Phi
_{(1,1),(1,1)}]  \nonumber \\
\lbrack \Phi _{(5,1),(5,1)}]\times [\Phi _{(1,1),(5,1)}] &=&[\Phi
_{(5,1),(1,1)}]  \nonumber \\
\lbrack \Phi _{(5,1),(5,1)}]\times [\Phi _{(5,1),(1,1)}] &=&[\Phi
_{(1,1),(5,1)}]  \nonumber
\end{eqnarray}

\section{\protect\medskip The correlation functions and the structure
constants}

In a 2-dimensionnal conformal field theory, the form of the two and 3-point
correlation functions are fixed only from symmetry considerations.
Considering only primary fields, and for a particular normalisation, the two
and three point functions are written as 
\begin{equation}
\left\langle \Phi _{h_{1},\overline{h}_{1}}(z_{1},\overline{z}_{1})\Phi
_{h_{2},\overline{h}_{2}}(z_{2},\overline{z}_{2})\right\rangle =\frac{%
(-1)^{s_{1}}\delta _{h_{1},h_{2}}\delta _{\overline{h}_{1},\overline{h}_{2}}%
}{z_{12}^{2h_{1}}\overline{z}_{12}^{2\overline{h}_{1}}}  \label{eq2-14}
\end{equation}

where $s_{1}=h_{1}-\overline{h}_{1}$ is the spin of the field $\Phi _{h_{1},%
\overline{h}_{1}}$, and

\begin{eqnarray}
&&\left\langle \Phi _{h_{1},\overline{h}_{1}}(z_{1},\overline{z}_{1})\Phi
_{h_{2},\overline{h}_{2}}(z_{2},\overline{z}_{2})\Phi _{h_{3},\overline{h}%
_{3}}(z_{3},\overline{z}_{3})\right\rangle  \label{eq2-15} \\
&=&\frac{C_{123}}{%
z_{12}^{h_{1}+h_{2}-h_{3}}z_{23}^{h_{2}+h_{3}-h_{1}}z_{13}^{h_{1}+h_{3}-h_{2}}%
}\times \frac{1}{(z_{ij}\rightarrow \overline{z}_{ij}\ ;\ h_{i}\rightarrow 
\overline{h}_{i})}  \nonumber
\end{eqnarray}
with $z_{ij}=z_{i}-z_{j}$. In the limit $z_{1}\rightarrow z_{2}$ (short
distance) the 3-point correlation behaves like

\begin{eqnarray}
&&\left\langle \Phi _{h_{1},\overline{h}_{1}}(z_{1},\overline{z}_{1})\Phi
_{h_{2},\overline{h}_{2}}(z_{2},\overline{z}_{2})\Phi _{h_{3},\overline{h}%
_{3}}(z_{3},\overline{z}_{3})\right\rangle  \nonumber \\
&\approx &\frac{\widetilde{C}_{123}}{z_{12}^{h_{1}+h_{2}-h_{3}}}\frac{1}{%
z_{23}^{2h_{3}}\ \ \times \ \ c.c}  \nonumber \\
&\approx &\frac{\widetilde{C}_{123}}{z_{12}^{h_{1}+h_{2}-h_{3}}\overline{z}%
_{12}^{\overline{h}_{1}+\overline{h}_{2}-\overline{h}_{3}}}\left\langle \Phi
_{h_{3},\overline{h}_{3}}(z_{2},\overline{z}_{2})\Phi _{h_{3},\overline{h}%
_{3}}(z_{3},\overline{z}_{3})\right\rangle  \label{eq2-16}
\end{eqnarray}
This means that the product $\Phi _{h_{1},\overline{h}_{1}}\times \Phi
_{h_{2},\overline{h}_{2}}$ contains the field $\Phi _{h_{3},\overline{h}%
_{3}} $ with strength $\widetilde{C}_{123}$\medskip . As already mentioned
the constants $C_{ijk}$ of the 3-point correlation function are then related
to the structure constants of the operator product algebra by 
\[
C_{123}=(-1)^{s_{3}}\widetilde{C}_{123} 
\]

The 4-point correlation functions are fixed up to an undermined factor
depending on the variable $z=\frac{z_{12}z_{34}}{z_{13}z_{24}}$ 
\begin{equation}
G(z_{1},...,\overline{z}_{4})=\left\langle \Phi _{1}\Phi _{2}\Phi _{3}\Phi
_{4}\right\rangle =f(z,\overline{z})\prod_{i<j}z_{ij}^{-h_{i}-h_{j}+h/3}%
\overline{z}_{ij}^{-\overline{h}_{i}-\overline{h}_{j}+\overline{h}/3}
\label{eq2-17}
\end{equation}
where $h=\sum_{i=1}^{4}$\medskip $h_{i}$ and $\overline{h}=\sum_{i=1}^{4}$%
\medskip $\overline{h}_{i}$ .

Using the operator product expansion (\ref{eq 2-9}), one can write $%
G(z_{1},...,\overline{z}_{4})$ in terms of the structure constants and
because of the fact that it is not obvious for which pairs of fields we
should compute the operator product first (the duality or the crossing
symmetry), one obtains strong constraints on the structure constants : the
bootstrap equations. In the $s$-channel ($z_{12},z_{34}\rightarrow 0,$or $%
z\rightarrow 0$ ) one obtains 
\begin{equation}
G(z_{1},...,\overline{z}_{4})=\sum_{m}\frac{(-1)^{s_{m}}\widetilde{C}_{12m}%
\widetilde{C}_{34m}\left[ 1+\mathcal{O}(z_{12},\overline{z}_{34})\right] }{%
z_{12}^{h_{1}+h_{2}-h_{m}}z_{34}^{h_{3}+h_{4}-h_{m}}z_{24}^{2h_{m}}\overline{%
z}_{12}^{\overline{h}_{1}+\overline{h}_{2}-\overline{h}_{m}}\overline{z}%
_{34}^{\overline{h}_{3}+\overline{h}_{4}-\overline{h}_{m}}\overline{z}%
_{24}^{2\overline{h}_{m}}}  \label{eq2-18}
\end{equation}
and in the $t$-channel ($z_{14},z_{23}\rightarrow 0,$or $(1-z)\rightarrow 0$
) 
\begin{equation}
G(z_{1},...,\overline{z}_{4})=\sum_{m}\frac{(-1)^{s_{m}}\widetilde{C}_{14m}%
\widetilde{C}_{23m}\left[ 1+\mathcal{O}(z_{14},\overline{z}_{23})\right] }{%
z_{41}^{h_{1}+h_{4}-h_{m}}z_{23}^{h_{3}+h_{2}-h_{m}}z_{13}^{2h_{m}}\overline{%
z}_{41}^{\overline{h}_{1}+\overline{h}_{4}-\overline{h}_{m}}\overline{z}%
_{23}^{\overline{h}_{3}+\overline{h}_{2}-\overline{h}_{m}}\overline{z}%
_{13}^{2\overline{h}_{m}}}  \label{eq2-19}
\end{equation}
On the other hand as one can always factorize the contribution of each
primary into an holomorphic part and an anti-holomorphic \medskip one, the
4-point correlation function can be written in the factorized forms 
\begin{equation}
G(z_{1},....,\overline{z}_{4})=\sum_{k,l}\ A_{kl}\ F_{k}(z_{1},..z_{4})\ 
\overline{F}_{l}(\overline{z}_{1},..,\overline{z}_{4})  \label{eq 2-20}
\end{equation}
where the $F_{i}$ and $\overline{F}_{i}$ are the conformal blocks. In the
case where all the fields present in the correlation are symmetric
left-right combination (i.e are spinless), the coefficients $A_{kl}$ are
diagonals. This is always the case in the models of the A-series but not for
those of the D-series because of the presence of spin fields.

\section{The conformal blocks and their integral representation}

To determine the structure constants for the $(D_{4},A_{6})$ model, we use
the method first used by Dotsenko in \cite{Dotsenko} for the models of the
A-series and later adapted by McCabe for the $D_{5}$ model in \cite{Mccabe1}
and \cite{Mccabe2}. It consists at first in finding the integral
representation of the conformal blocks in the coulomb gas formulation and to
write their approximate forms \ as well as that of the associated 4-point
correlation function (\ref{eq 2-20}) in both the $s$- and $t$- channel.
After that one can compare the obtained expressions for $G(z_{1},....,%
\overline{z}_{4})$ with those given by the equations (\ref{eq2-18}) and (\ref
{eq2-19}) respectively and extract the structure constants.

In this section we try to reproduce, with very few details and using the
same notation as in \cite{Mccabe1}, the most important steps in the
determination of the integral representation of the conformal blocks for the
special case of the 4-point correlations containing the field $\Phi
_{(3,1),(3,1)}$ of the form 
\begin{equation}
G(z_{1},....,\overline{z}_{4})=\left\langle \Phi _{(m,n),(\overline{m},%
\overline{n})}\Phi _{(3,1),(3,1)}\Phi _{(3,1),(3,1)}\Phi _{(m,n),(\overline{m%
},\overline{n})}\right\rangle  \label{eq 2-20'}
\end{equation}
In the coulomb gas approach, the conformal blocks are written as 
\begin{equation}
F_{k}(z_{i})=\left\langle \prod_{i=1}^{3}V_{\alpha
_{r_{i}s_{i}}}(z_{i})V_{-\alpha _{r_{4}s_{4}}+2\alpha
_{0}}(z_{4})Q_{+}^{N}Q_{-}^{M}\right\rangle  \label{eq2-21}
\end{equation}
\medskip where the $V_{\alpha _{rs}}(z)$ is a vertex operator of charge $%
\alpha _{rs}$ 
\[
\alpha _{rs}=-\frac{1}{2}([r-1]\alpha _{+}+[s-1]\alpha _{-}) 
\]
and conformal weight 
\begin{eqnarray*}
h_{rs} &=&\alpha _{rs}(\alpha _{rs}-2\alpha _{0}) \\
2\alpha _{0} &=&\alpha _{+}+\alpha _{-}\;\;\;\;\;;\;\;\;\;\;\alpha
_{+}\alpha _{-}=-1
\end{eqnarray*}
with 
\[
\alpha _{+}=\sqrt{\frac{m+1}{m}}=\sqrt{\frac{7}{6}}\;\;\;\;\;\;;\;\;\;\;\;\;%
\alpha _{-}=-\sqrt{\frac{m}{m+1}}=-\sqrt{\frac{6}{7}}\; 
\]
The operators $Q_{\pm }$ are the screen operators, they are integrals over
closed contour of the vertex $V_{\alpha _{\pm }}$ of conformal dimension 1 
\[
Q_{\pm }=\oint_{\widetilde{C}}duV_{\alpha _{\pm }}(u) 
\]
and their exponents $N$ and $M$ are fixed by the neutrality condition 
\[
\sum_{i=1}^{3}\alpha _{r_{i}s_{i}}-\alpha _{r_{4}s_{4}}+2\alpha _{0}+N\alpha
_{+}+M\alpha _{-}=0 
\]
For the chosen 4-point correlation (\ref{eq 2-20'}), the last equation is
solved for $N=0$ and $M=2$. Noting that 
\[
\left\langle \prod_{i=1}^{K}V_{\alpha _{i}}(z_{i})\right\rangle =\left\{ 
\begin{array}{c}
\prod_{i<j}^{K}(z_{i}-z_{j})^{2\alpha _{i}\alpha _{j}}\;\;\;\;\;\text{if \ \
\ \ \ \ \ \ }\sum_{i=1}^{K}\alpha _{i}=2\alpha _{0} \\ 
\\ 
0\;\ \ \ \ \ \ \ \ \ \ \ \ \ \ \ \ \ \ \ \ \ \ \ \text{otherwise}
\end{array}
\right\} 
\]
the conformal block is then written as 
\begin{eqnarray}
F_{k}(z_{i}) &=&(z_{12}z_{13})^{2\alpha _{mn}\alpha _{31}}(z_{23})^{2\alpha
_{31}^{2}}(z_{24}z_{34})^{-2\alpha _{31}(\alpha _{mn}-\alpha _{+}-\alpha
_{-})}(z_{14})^{-2\alpha _{mn}(\alpha _{mn}-\alpha _{+}-\alpha _{-})} 
\nonumber \\
&&\times \oint_{\widetilde{C}_{1}(k)}dv_{1}\oint_{\widetilde{C}%
_{2}(k)}dv_{2}\left[ \left( v_{1}-z_{1}\right) \left( v_{2}-z_{1}\right)
\right] ^{2\alpha _{-}\alpha _{mn}}  \nonumber \\
&&\qquad \left[ \left( v_{1}-z_{2}\right) \left( v_{1}-z_{3}\right) \left(
v_{2}-z_{2}\right) \left( v_{2}-z_{3}\right) \right] ^{^{2\alpha _{-}\alpha
_{31}}}  \nonumber \\
&&\qquad \left[ \left( v_{1}-z_{4}\right) \left( v_{2}-z_{4}\right) \right]
^{-2\alpha _{-}(\alpha _{mn}-\alpha _{+}-\alpha _{-})}(v_{1}-v_{2})^{2\alpha
_{-}^{2}}  \label{eq2-22}
\end{eqnarray}
A simpler form of this equation is given by 
\begin{eqnarray}
F_{k}(z_{i}) &=&\prod_{i=1}^{4}\left( \frac{d\omega }{dz_{i}}\right)
^{h_{i}}F_{k}(0,z,1,\infty )  \nonumber \\
&=&(z_{14})^{-2h_{mn}}\left( \frac{z_{14}}{z_{31}z_{24}}\right)
^{2h_{31}}z^{2\alpha _{mn}\alpha _{31}}(1-z)^{2\alpha _{31}^{2}}  \nonumber
\\
&&\times I_{k}(2\alpha _{mn}\alpha _{-},2\alpha _{31}\alpha _{-},2\alpha
_{31}\alpha _{-},2\alpha _{-}^{2},z)  \label{eq2-23}
\end{eqnarray}
with 
\begin{equation}
I_{k}(a,b,\widetilde{C},g,z)=\oint_{\widetilde{C}_{1}(k)}dv_{1}\oint_{%
\widetilde{C}_{2}(k)}dv_{2}\ f(v_{1},v_{2};a,b,\widetilde{C},g,z)
\label{eq 2-25}
\end{equation}

\begin{equation}
f(v_{1},v_{2};a,b,\widetilde{C},g,z)=\left[ v_{1}v_{2}\right] ^{a}\left[
\left( v_{1}-1\right) \left( v_{2}-1\right) \right] ^{b}\left[ \left(
v_{1}-z\right) \left( v_{2}-z\right) \right] ^{\widetilde{C}%
}(v_{1}-v_{2})^{g}  \label{eq 2-26'}
\end{equation}
To obtain (\ref{eq2-23}) one perform the global conformal transformation 
\[
z\rightarrow \omega (z)=\frac{z_{34}}{z_{31}}\left( \frac{z-z_{1}}{z-z_{4}}%
\right) 
\]
which fixes $z_{1},z_{3}$ and $z_{4}$ to 0,1 and $\infty $ respectively,
while $z_{2}$ is transformed into $z=\frac{z_{12}z_{34}}{z_{13}z_{24}}$ .

There is in principle many possibilities for the choice of the integration
contours appearing in (\ref{eq2-23}). But because of the fact that the
integrand has branch cuts at $0,1,z$ and $\infty $, it is shown then that
there is only three independants integration contours (see \cite{Dotsenko}
for more details) leading to three different blocks defined with the
following integrals 
\begin{mathletters}
\begin{eqnarray}
I_{1}(a,b,c,g,z)
&=&\int_{0}^{z}dv_{1}\int_{0}^{v_{1}}dv_{2}f(v_{1},v_{2};a,b,c,g,z)
\label{eq 2-27a} \\
I_{2}(a,b,c,g,z) &=&\int_{1}^{\infty
}dv_{1}\int_{1}^{v_{1}}dv_{2}f(v_{1},v_{2};a,b,c,g,z)  \label{eq 2-27b} \\
I_{3}(a,b,c,g,z) &=&\int_{1}^{\infty
}dv_{1}\int_{0}^{z}dv_{2}f(v_{1},v_{2};a,b,c,g,z)  \label{eq 2-27c}
\end{eqnarray}
In the $s$-channel ($z\rightarrow 0)$ they behave like 
\end{mathletters}
\begin{mathletters}
\begin{eqnarray}
I_{1}(a,b,c,g,z) &\simeq &N_{1}[1+o(z)]  \label{eq 2-28a} \\
I_{2}(a,b,c,g,z) &\simeq &N_{2}z^{1+a+c}[1+o(z)]  \label{eq 2-28b} \\
I_{3}(a,b,c,g,z) &\simeq &N_{3}z^{2(1+a+c)+g}[1+o(z)]  \label{eq 2-28c}
\end{eqnarray}
Now one can write the 4-point correlation function of interest (\ref{eq
2-20'}) in the $s$-channel and compare it with the corresponding one given
by (\ref{eq2-18}) to obtain some equations involving the desired structure
constants. To do the same in the $t$-channel one uses the monodromy
properties of the integrals $I_{k}$ to transforme them into functions of $%
(1-z)$ with the monodromy matrices 
\end{mathletters}
\[
I_{k}(a,b,c,g,z)=\sum_{j}\gamma _{kj}I_{j}(b,a,c,g,1-z) 
\]
$\left[ \gamma \right] $ is the monodromy matrix. It's general expression as
well that of the $N_{k}$ are given in \cite{Dotsenko}.

In the particular case where the correlation function studied contains an
operator having a null vector at the level one, like $\Phi _{(1,1),(%
\overline{m},\overline{n})}(z,\overline{z})$ for example, we have the
additional constraint 
\[
\partial _{z}\Phi _{(1,1),(\overline{m},\overline{n})}(z,\overline{z})=0 
\]

This equation with the usual ones, obtained from the global conformal
invariance, permit to obtain a very simple forms of the conformal blocks 
\cite{Mccabe1}.

\section{Application}

As an application of the previous algorithm, in this section, explicit
values of some structure constants of the subalgebra will be calculated.

\subsection{The structure constant $\widetilde{C}_{(51\mid 11),(11\mid
51),(51\mid 51)}:$}

It is obtained by considering the following correlation function

\begin{equation}
G_{1}=\left\langle \Phi _{(11\mid 51)}(1)\Phi _{(51\mid 11)}(2)\Phi
_{(51\mid 11)}(3)\Phi _{(11\mid 51)}(4)\right\rangle
\end{equation}
From the bootstrap equations (\ref{eq2-18})and (\ref{eq2-19}) one can write
in the $s-$channel : 
\begin{equation}
G_{1}=\frac{\left( \widetilde{C}_{(51\mid 11),(11\mid 51),(51\mid
51)}\right) ^{2}\left[ 1+\mathcal{O}(z,\overline{z})\right] }{\left( z_{24}%
\overline{z}_{24}\right) ^{10}}
\end{equation}
and in the $t-$channel : 
\begin{equation}
G_{1}=\frac{\left[ 1+\mathcal{O}(1-z,1-\overline{z})\right] }{\left( z_{23}%
\overline{z}_{41}\right) ^{10}}
\end{equation}
On the other hand from the integral representation, one finds only one block
and can write

\begin{equation}
G_{1}=\frac{A}{\left( z_{23}\overline{z}_{41}\right) ^{10}}
\end{equation}
when $z\rightarrow 0$ the last expression behaves like

\begin{equation}
G_{1}=\frac{A\left[ 1+\mathcal{O}(z,\overline{z})\right] }{\left( z_{24}%
\overline{z}_{24}\right) ^{10}}
\end{equation}
and for ($1-z$) $\rightarrow 0$

\begin{equation}
G_{1}=\frac{A}{\left( z_{23}\overline{z}_{41}\right) ^{10}}
\end{equation}
Comparing the leading terms from the correponding expressions one find :

\begin{equation}
(\widetilde{C}_{(51\mid 11),(11\mid 51),(51\mid 51)})^{2}=1
\end{equation}

\subsection{The structure constant $\widetilde{C}_{(51\mid 11)+-}:$}

Here we use the function

\begin{equation}
G_{2}=\left\langle \Phi _{(51\mid 11)}(1)\Phi ^{+}(2)\Phi ^{+}(3)\Phi
_{(51\mid 11)}(4)\right\rangle
\end{equation}
The bootstrap equations give, in the $s-$channel : 
\begin{equation}
G_{2}=\frac{-\left( \widetilde{C}_{(51\mid 11)+-}\right) ^{2}\left[ 1+%
\mathcal{O}(z,\overline{z})\right] }{\left( z_{12}z_{34}\right) ^{5}\left(
z_{24}\overline{z}_{24}\right) ^{\frac{8}{3}}}
\end{equation}
and in the $t-$channel :

\begin{equation}
G_{2}=\frac{-\left[ 1+\mathcal{O}(1-z,1-\overline{z})\right] }{%
(z_{41})^{10}\left( z_{23}\overline{z}_{23}\right) ^{\frac{8}{3}}}
\end{equation}
The development in terms of the conformal blocks is given by

\begin{equation}
G_{2}=\sum_{k=1}^{3}-A_{k}\frac{z^{-5}(1-z)^{\frac{7}{3}}}{z_{14}^{\frac{22}{%
3}}(\overline{z}_{23}z_{31}z_{24})^{\frac{8}{3}}}I_{k}\left( 5,\frac{-7}{3},%
\frac{-7}{3},\frac{7}{3};z\right)
\end{equation}
\ The blocks $\overline{F}_{l}(\overline{z}_{1},..,\overline{z}_{4})$ are
determined using the fact that the correlation $G_{2}\ $involves null state
vectors at level one. From the equations (\ref{eq 2-28a}),(\ref{eq 2-28b})
and (\ref{eq 2-28c}) for the behaviour of the $I_{k}$'s when $z\rightarrow 0$%
, we have

\begin{equation}
G_{2}=\frac{-z^{-5}\left[ 1+\mathcal{O}(z,\overline{z})\right] }{z_{14}^{%
\frac{22}{3}}(\overline{z}_{23}z_{31}z_{24})^{\frac{8}{3}}}\left(
A_{1}N_{1}+A_{2}N_{2}z^{\frac{11}{3}}+A_{3}N_{3}z^{\frac{29}{3}}\right)
\end{equation}
and a comparison of the both expressions in the $s-$channel gives :

\begin{equation}
-\left( \widetilde{C}_{(51\mid 11)+-}\right) ^{2}=-A_{1}N_{1}\qquad ;\qquad
A_{2}=A_{3}=0
\end{equation}
while the same thing when $1-z\rightarrow 0$ yields to 
\begin{eqnarray}
G_{2} &=&\frac{-(1-z)^{\frac{7}{3}}\left[ 1+\mathcal{O}(1-z,1-\overline{z}%
)\right] }{z_{14}^{\frac{22}{3}}(\overline{z}_{23}z_{31}z_{24})^{\frac{8}{3}}%
}\times \{A_{1}\gamma _{11}I_{1}\left( \frac{-7}{3},5,\frac{-7}{3},\frac{7}{3%
};1-z\right)  \nonumber \\
&&+A_{1}\gamma _{12}I_{2}\left( \frac{-7}{3},5,\frac{-7}{3},\frac{7}{3}%
;1-z\right) +A_{1}\gamma _{13}I_{3}\left( \frac{-7}{3},5,\frac{-7}{3},\frac{7%
}{3};1-z\right) \}  \nonumber \\
&=&-\frac{1}{2}A_{1}N_{3}\frac{(1-z)^{\frac{7}{3}}(1-z)^{-5}\left[ 1+%
\mathcal{O}(1-z,1-\overline{z})\right] }{z_{14}^{\frac{22}{3}}(\overline{z}%
_{23}z_{31}z_{24})^{\frac{8}{3}}}
\end{eqnarray}

\begin{equation}
-1=-\frac{1}{2}A_{1}N_{3}
\end{equation}
Finally from the comparison one obtains

\begin{equation}
\left( \widetilde{C}_{(51\mid 11)+-}\right) ^{2}=2\frac{N_{1}\left( 5,\frac{%
-7}{3},\frac{-7}{3},\frac{7}{3}\right) }{N_{3}\left( \frac{-7}{3},5,\frac{-7%
}{3},\frac{7}{3}\right) }
\end{equation}

\subsection{The structure constant $\widetilde{C}_{+++}:$}

Considering the function

\begin{equation}
G_{3}=\left\langle \Phi ^{+}(1)\Phi ^{+}(2)\Phi ^{+}(3)\Phi
^{+}(4)\right\rangle
\end{equation}
we have in the $s-$channel :

\begin{equation}
G_{3}=\left\{ \frac{1}{\left| z_{12}z_{34}\right| ^{\frac{16}{3}}}+\frac{%
\left( \widetilde{C}_{+++}\right) ^{2}}{\left| z_{12}z_{34}\right| ^{\frac{8%
}{3}}\left( z_{24}\overline{z}_{24}\right) ^{\frac{8}{3}}}+\frac{\left( 
\widetilde{C}_{++(51\mid 51)}\right) ^{2}}{\left| z_{12}z_{34}\right| ^{%
\frac{-14}{3}}\left( z_{24}\overline{z}_{24}\right) ^{10}}\right\} \left[ 1+%
\mathcal{O}(z,\overline{z})\right]  \label{c+++z0}
\end{equation}
and in the $t-$channel :

\begin{equation}
G_{3}=\left\{ \frac{1}{\left| z_{41}z_{23}\right| ^{\frac{16}{3}}}+\frac{%
\left( \widetilde{C}_{+++}\right) ^{2}}{\left| z_{41}z_{23}\right| ^{\frac{8%
}{3}}\left( z_{13}\overline{z}_{13}\right) ^{\frac{8}{3}}}+\frac{\left( 
\widetilde{C}_{++(51\mid 51)}\right) ^{2}}{\left| z_{41}z_{23}\right| ^{%
\frac{-14}{3}}\left( z_{31}\overline{z}_{31}\right) ^{10}}\right\} \left[ 1+%
\mathcal{O}(1-z,1-\overline{z})\right]
\end{equation}
The factorized form is

\begin{equation}
G_{3}=\sum_{k,l=1}^{3}A_{kl}\frac{\left| z\left( 1-z\right) \right| ^{\frac{%
14}{3}}}{\left| z_{31}z_{24}\right| ^{\frac{16}{3}}}I_{k}\left( \frac{-7}{3},%
\frac{-7}{3},\frac{-7}{3},\frac{7}{3};z\right) \times \overline{I}_{l}\left( 
\frac{-7}{3},\frac{-7}{3},\frac{-7}{3},\frac{7}{3};\overline{z}\right)
\label{eq-G3}
\end{equation}
The comparison between its behavior when $z\rightarrow 0$ and (\ref{c+++z0})
gives

\begin{eqnarray}
A_{33}N_{3}^{2} &=&1 \\
A_{11}N_{1}^{2} &=&\left( \widetilde{C}_{++(51\mid 51)}\right) ^{2} \\
A_{22}N_{2}^{2} &=&\left( \widetilde{C}_{+++}\right) ^{2}
\end{eqnarray}
while that when ($1-z)\rightarrow 0$ leads to

\begin{eqnarray}
1 &=&\frac{N_{3}^{2}}{4}\left( A_{11}+A_{22}+A_{33}\right) \\
\left( \widetilde{C}_{+++}\right) ^{2} &=&N_{2}^{2}\left(
A_{11}+A_{33}\right) \\
\left( \widetilde{C}_{++(51\mid 51)}\right) ^{2} &=&\frac{N_{1}^{2}}{4}%
\left( A_{11}+A_{22}+A_{33}\right)
\end{eqnarray}
Resolving the two systems of equations , we find :

\begin{equation}
\left( \widetilde{C}_{+++}\right) ^{2}=8\frac{N_{2}^{2}\left( \frac{-7}{3},%
\frac{-7}{3},\frac{-7}{3},\frac{7}{3}\right) }{N_{1}^{2}\left( \frac{-7}{3},%
\frac{-7}{3},\frac{-7}{3},\frac{7}{3}\right) }\frac{N_{1}^{2}\left( 5,\frac{%
-7}{3},\frac{-7}{3},\frac{7}{3}\right) }{N_{3}^{2}\left( \frac{-7}{3},5,%
\frac{-7}{3},\frac{7}{3}\right) }
\end{equation}

It is noted here that $G_{3}$ can be calculated as a correlator of the A-
series, however we calculated it with non diagonal blocks form and we
observed that only non vanishing elements of $A_{ij}$\ are those of the
diagonal, this fact is not trivial when considering $G_{3}$\ with different
action of $Z_{2}$.

A simillar calculus with $\left\langle \Phi ^{-}(1)\Phi ^{-}(2)\Phi
^{-}(3)\Phi ^{-}(4)\right\rangle $ gives,

\[
\left( \widetilde{C}_{--+}\right) ^{2}=\left( \widetilde{C}_{+++}\right)
^{2} 
\]

\subsection{The sign of the structure constants :}

\subsubsection{sign of $\widetilde{C}_{+--}:$}

Let consider the correlation:

\[
G_{4}=\left\langle \Phi ^{+}(1)\Phi ^{+}(2)\Phi ^{-}(3)\Phi
^{-}(4)\right\rangle 
\]

In the $s$- and $t$-channel we have respectively

\begin{equation}
G_{4}=\left\{ \frac{1}{\left| z_{12}z_{34}\right| ^{\frac{16}{3}}}+\frac{%
\widetilde{C}_{+++}\widetilde{C}_{--+}}{\left| z_{12}z_{34}\right| ^{\frac{8%
}{3}}\left( z_{24}\overline{z}_{24}\right) ^{\frac{8}{3}}}+\frac{\widetilde{C%
}_{++(51\mid 51)}\widetilde{C}_{--(51\mid 51)}}{\left| z_{12}z_{34}\right| ^{%
\frac{-14}{3}}\left( z_{24}\overline{z}_{24}\right) ^{10}}\right\} \left[ 1+%
\mathcal{O}(z,\overline{z})\right]
\end{equation}

\begin{equation}
G_{4}=\left\{ \frac{\left( \widetilde{C}_{+--}\right) ^{2}}{\left|
z_{41}z_{23}\right| ^{\frac{8}{3}}\left( z_{13}\overline{z}_{13}\right) ^{%
\frac{8}{3}}}+\frac{-\left( \widetilde{C}_{+-(51\mid 11)}\right) ^{2}}{%
\left( z_{41}z_{23}\right) ^{\frac{-7}{3}}\left( \overline{z}_{41}\overline{z%
}_{23}\right) ^{\frac{8}{3}}z_{13}^{10}}+\frac{-\left( \widetilde{C}%
_{+-(11\mid 51)}\right) ^{2}}{\left( z_{41}z_{23}\right) ^{\frac{8}{3}%
}\left( \overline{z}_{41}\overline{z}_{23}\right) ^{\frac{-7}{3}}\overline{z}%
_{13}^{10}}\right\} \left[ 1+\mathcal{O}(1-z,1-\overline{z})\right]
\end{equation}

A comparison with (\ref{eq-G3}) in the $\mathit{s}$-channel leads to,

\begin{eqnarray}
A_{22}N_{2}^{2} &=&1  \label{second1} \\
A_{31}N_{1}N_{3} &=&\widetilde{C}_{+++}\widetilde{C}_{--+}  \nonumber \\
A_{13}N_{1}N_{3} &=&\widetilde{C}_{++(51\mid 51)}\widetilde{C}_{--(51\mid
51)}  \nonumber
\end{eqnarray}

where, 
\[
A_{13}=A_{31} 
\]

and in the t-channel to\textit{,}

\begin{eqnarray}
\left( A_{13}+\frac{A_{22}}{2}\right) \frac{N_{3}^{2}}{2} &=&\left( 
\widetilde{C}_{+--}\right) ^{2}  \label{second} \\
A_{13}N_{2}^{2} &=&-\left( \widetilde{C}_{+-(51\mid 11)}\right) ^{2} 
\nonumber \\
\left( A_{13}+\frac{A_{22}}{2}\right) \frac{N_{1}^{2}}{2} &=&-\left( 
\widetilde{C}_{+-(11\mid 51)}\right) ^{2}  \nonumber
\end{eqnarray}

The second equation in (\ref{second}) gives $A_{13}\prec 0$\ , and the
second one of (\ref{second1}) leads to

\[
\widetilde{C}_{--+}=-\widetilde{C}_{+++} 
\]

Finally, we note from the results obtained from $G_{4}$\ that if we choose $%
\widetilde{C}_{+++}$\ to be positive (as all the structure constants of the
A-series) $\widetilde{C}_{--+}$ will be of negative sign.

\subsubsection{Sign of $\widetilde{C}_{--(51\mid 51)}$}

\textit{\ }From the correlation function

\begin{equation}
G_{5}=\left\langle \Phi ^{-}(1)\Phi ^{-}(2)\Phi ^{+}(3)\Phi _{(51\mid
51)}(4)\right\rangle
\end{equation}

we have

\begin{equation}
\widetilde{C}_{--+}\widetilde{C}_{++(51\mid 51)}=\widetilde{C}_{-+-}%
\widetilde{C}_{--(51\mid 51)}
\end{equation}

and if we take $\widetilde{C}_{++(51\mid 51)}$ to be positive we obtain

\begin{equation}
\widetilde{C}_{++(51\mid 51)}=\widetilde{C}_{--(51\mid 51)}
\end{equation}

It is then noted here that the calculus of the conformal blocks and the
corresponding monodromy matrices for the correlation of type $G_{5}$ are
obtained in the same way as was done for the correlation (\ref{eq 2-20'}) .

\subsubsection{Sign of $\widetilde{C}_{(51\mid 11)(11\mid 51)(51\mid 51)}$}

Consider the correlation

\begin{equation}
G_{5}=\left\langle \Phi _{(51\mid 11)}(1)\Phi ^{+}(2)\Phi ^{+}(3)\Phi
_{(11\mid 51)}(4)\right\rangle
\end{equation}
\ 

the comparison gives

\begin{equation}
-\widetilde{C}_{(51\mid 11)+-}\widetilde{C}_{(11\mid 51)+-}=\widetilde{C}%
_{(51\mid 11),(11\mid 51),(51\mid 51)}\widetilde{C}_{++(51\mid 51)}
\end{equation}

The constants structure in l.h.s of the relation above can be choosen to be
positives ( this not altere the form of the two points correlation). This
gives $\widetilde{C}_{(51\mid 11),(11\mid 51),(51\mid 51)}$

\[
\widetilde{C}_{(51\mid 11),(11\mid 51),(51\mid 51)}=-1 
\]

The values for the remaining structure constants are of course obtained in
the same manner. They are resumed in the table 2.

\medskip 
\begin{table}[tbp] \centering%
\begin{tabular}{|c|c|}
\hline\hline
\multicolumn{1}{||c|}{\textbf{The structure Constants}} & 
\multicolumn{1}{||c||}{\textbf{Value}} \\ \hline
$\widetilde{C}_{+--}$ & $\frac{-25}{18}\frac{\sqrt{6}\Gamma \left( \frac{2}{3%
}\right) ^{6}}{\pi ^{3/2}\Gamma \left( \frac{5}{6}\right) ^{3}}$ \\ \hline
$\widetilde{C}_{+++}$ & $\frac{25}{18}\frac{\sqrt{6}\Gamma \left( \frac{2}{3}%
\right) ^{6}}{\pi ^{3/2}\Gamma \left( \frac{5}{6}\right) ^{3}}$ \\ \hline
$\widetilde{C}_{(51\mid 11)+-}=\widetilde{C}_{(11\mid 51)+-}$ & $\frac{2%
\sqrt{1173}}{81}$ \\ \hline
$\widetilde{C}_{++(51\mid 51)}=\widetilde{C}_{--(51\mid 51)}$ & $\frac{1564}{%
2187}$ \\ \hline
$\widetilde{C}_{(51\mid 11),(11\mid 51),(51\mid 51)}$ & $-1$ \\ \hline
\end{tabular}
\caption{The structure constants\label{key}}%
\end{table}%

\section{Summary}

To conclude this work, we would like to notice that the use of the
transformation \textbf{T}$_{2}$ as the $Z_{2}$-symmetry in our case doesn't
give any inconsistencies with the non chiral fusion rules and the bootstrap
equations.  We also notice that the relative signs of the structure
constants we obtained are in concordance with those obtained in \cite
{petkova}.

\textbf{\ }

\end{document}